\documentstyle[11pt,aaspp4,flushrt]{article}

\def\longpagen[#1]{\setlength{\textheight}{#1}}

\newcommand\gsim{\mathrel{\raise.3ex\hbox{$>$}\mkern-14mu
             \lower0.6ex\hbox{$\sim$}}}
\newcommand\lsim{\mathrel{\raise.3ex\hbox{$<$}\mkern-14mu
             \lower0.6ex\hbox{$\sim$}}}

\def\arcs{\ifmmode {^{\scriptscriptstyle\prime\prime}}
          \else $^{\scriptscriptstyle\prime\prime}$\fi}
\def\arcm{\ifmmode {^{\scriptscriptstyle\prime}}
          \else $^{\scriptscriptstyle\prime}$\fi}
\newdimen\sa  \newdimen\sb
\def\parcs{\sa=.07em \sb=.03em
     \ifmmode $\rlap{.}$^{\scriptscriptstyle\prime\kern -\sb\prime}$\kern -\sa$
     \else \rlap{.}$^{\scriptscriptstyle\prime\kern -\sb\prime}$\kern -\sa\fi}
\def\parcm{\sa=.08em \sb=.03em
     \ifmmode $\rlap{.}\kern\sa$^{\scriptscriptstyle\prime}$\kern-\sb$
     \else \rlap{.}\kern\sa$^{\scriptscriptstyle\prime}$\kern-\sb\fi}

\begin{document}

\title{DYNAMICS OF STARS AND GLOBULAR CLUSTERS IN M87}
\author{Aaron J. Romanowsky}
\affil{
Kapteyn Astronomical Institute, P.O. Box 800, 9700 AV Groningen, The Netherlands\\
and Harvard-Smithsonian Center for Astrophysics,
       60 Garden Street,
       Cambridge MA 02138 \\
       Email: {\tt romanow@astro.rug.nl}}
\author{Christopher S. Kochanek}
\affil{
Harvard-Smithsonian Center for Astrophysics,
       MS-51, 60 Garden Street,
       Cambridge MA 02138 \\
       Email: {\tt ckochanek@cfa.harvard.edu}}

\begin{abstract}
We examine the dynamics of the stars and globular clusters in the
nearby giant elliptical galaxy M87
and constrain the mass distribution,
using all the available data 
over a large range of radii, including higher-order moments of the stellar
line-of-sight velocity distributions
and the discrete velocities of over two hundred globular clusters.
We introduce an extension of spherical orbit modeling methods
that makes full use of all the information in the data,
and provides very robust constraints on the mass models.
We conclusively rule out a constant mass-to-light ratio model,
and infer that the radial density profile of the galaxy's dark halo
falls off more slowly than $r^{-2}$,
suggesting that the potential of the Virgo Cluster is already dominant at
$r \sim 300\arcs \sim 20$ kpc.

\end{abstract}
\keywords{
galaxies: elliptical and lenticular, cD ---
galaxies: halos ---
galaxies: individual (M87) ---
galaxies: kinematics and dynamics ---
galaxies: star clusters ---
galaxies: structure ---
globular clusters: general
}

\section{INTRODUCTION}

Elliptical galaxies
lack the dynamical simplicity of spiral galaxies,
posing well-known challenges for determining
their intrinsic properties
({\it e.g.}, mass distribution, shape, orbit structure).
The projected stellar velocity dispersion radial profile,
$\hat{\sigma}_{\rm p}(R)$,
is easily measured, but
is subject to a degeneracy between the mass distribution and the orbital anisotropy,
where radial variations in the stellar orbit types 
mask or mimic changes in the mass-to-light ratio ($M/L$)
(Binney \& Mamon 1982; Tonry 1983).
Crucial for resolving this degeneracy is the measurement of
higher-order moments of the stellar line-of-sight velocity distributions
(LOSVDs).
Such LOSVD information has been used to constrain the masses of central
supermassive black holes (van der Marel {\it et al.} 1998; Emsellem, Dejonghe, \& Bacon 1999; 
Cretton \& van den Bosch 1999;
Gebhardt {\it et al.} 2000) 
and dark halos (Carollo {\it et al.} 1995;
Rix {\it et al.} 1997; Gerhard {\it et al.} 1998;
Saglia {\it et al.} 2000;
Kronawitter {\it et al.} 2000).
Unfortunately, 
the surface brightness of an elliptical declines rapidly with radius,
making these measurements increasingly difficult at the large radii
where the
dark matter problem becomes most interesting
($\gsim$ 2 effective radii $R_{\rm eff}$).
Other techniques are necessary to probe this outer mass distribution.

In nearby galaxies,
the two main candidates for such probes are X-ray emission
from extended hot gas (see Fabbiano 1989 \S 4.4 for a review),
and the discrete velocities of either planetary nebulae
({\it e.g.}, Ciardullo, Jacoby, \& Dejonghe 1993;
Tremblay, Merritt, \& Williams 1995;
Hui {\it et al.} 1995; Arnaboldi {\it et al.} 1998) 
or globular clusters ({\it e.g.}, Cohen \& Ryzhov 1997, hereafter CR;
Kissler-Patig {\it et al.} 1999; Zepf {\it et al.} 2000).
However, the analysis of discrete kinematical measurements calls for more refined
dynamical models than either simple virial estimators
({\it e.g.}, Heisler, Tremaine, \& Bahcall 1985; Kent 1990; Haller \& Melia 1996)
or binning the data and using the Jeans equations
({\it e.g.},  
Federici {\it et al.} 1993;
Grillmair {\it et al.} 1994;
Tremblay {\it et al.} 1995).
These procedures do not guarantee a physical solution, make
unwarranted assumptions about the orbital anisotropy,
and do not make full use of the constraints provided by the data.
The first two deficiencies are alleviated by techniques
employing distribution function (DF) basis sets that permit the
full freedom of orbit types
({\it e.g.}, Merritt, Meylan, \& Mayor 1997; Mathieu \& Dejonghe 1999;
Saglia {\it et al.} 2000).
But these techniques typically bin the data in radius and velocity,
destroying potentially useful information,
such as the maximum observed velocity as a 
direct constraint on the escape velocity.
More general methods are needed to make better use of the information
contained in the discrete velocity data;
Merritt \& Saha (1993) and Merritt (1993) have demonstrated one such
approach using DF basis sets and maximum likelihood methods.
We develop a general dynamical method based on orbit modeling,
and apply it to a real galaxy, M87.

The giant central Virgo Cluster galaxy M87 (= NGC 4486),
although not a typical elliptical,
is nevertheless a prime candidate for modeling
because of the abundance of data available on it.
In addition to high-precision measurements of its
higher-order stellar velocity moments
and of the radially-extended velocity dispersion,
it has the largest available sample of globular cluster (GC) velocity measurements.
The stellar velocity measurements extend out to $\sim 1.5 R_{\rm eff}$,
and the GC measurements to $\sim 5 R_{\rm eff}$,
probing well into the area where the dynamics are
presumably dominated by the dark halo of the galaxy or of the Virgo Cluster.
The combination of these kinematical constraints
should provide robust limits on the dark matter distribution,
and on the orbital structures of both the stars and the GCs.

While a variety of modeling techniques have been used to study the stellar
dynamics of M87
(Sargent {\it et al.} 1978; Duncan \& Wheeler 1980; Binney \& Mamon 1982;
Newton \& Binney 1984; Richstone \& Tremaine 1985;
Dressler \& Richstone 1990; Tenjes, Einasto, \& Haud 1991;
van der Marel 1994; Merritt \& Oh 1997),
they concentrated on its central regions in order to constrain the mass
of a central black hole.
The stellar dynamics of its {\it halo} have not been modeled in detail, nor have
the higher-order velocity moments been used as strict constraints.
But the best available tracers of the halo mass distribution are 
ultimately the GCs and the hot gas,
which can be easily observed at large radii.
Some simple dynamical methods have been used to
estimate the halo mass using the GCs (Merritt \& Tremblay 1993; CR). 
They claim that the radially-rising velocity dispersion profile rules out
a constant $M/L$ model,
and that it
suggests that the GCs are in fact tracing the Virgo Cluster potential
rather than M87's own potential
(as suggested also by analyses of the X-ray emission and the
kinematics of the cluster galaxies;
{\it e.g.}, see Nulsen \& B\"{o}hringer 1995; McLaughlin 1999).
But a more rigorous dynamical model, allowing for the systematic orbital
uncertainties, is necessary to confirm these findings.
We will apply our general orbit modeling method, using a large sample
($\sim 200$) of discrete GC velocities,
to see how the GCs --- alone and jointly with the stellar dynamics ---
constrain 
the mass distribution.
In \S 2 we detail the observational constraints for M87.
We describe our modeling methods in \S 3,
\S 4 presents the results, and the conclusions are in \S 5.

\section{OBSERVATIONAL CONSTRAINTS}

Here we summarize
the observational constraints on M87.
These include the stellar surface brightness 
(\S 2.1);
velocity measurements of the stars 
(\S 2.2);
the globular cluster surface density 
(\S 2.3);
and globular cluster discrete velocity measurements 
(\S 2.4).
We assume a distance to M87 of 15 Mpc (Pierce {\it et al.} 1994), 
so that $1\arcs=73$ pc and $1\arcm=4.4$ kpc.
The stellar effective radius is $R_{\rm eff}\simeq 100\arcs$ 
(Burstein {\it et al.} 1987; 
Zeilinger, M{\o}ller, \& Stiavelli 1993).

\subsection{Stellar Surface Brightness}

For the stellar surface brightness radial profile $\mu(R)$ in the core region
($R$=0\parcs01--8\arcs)
we use the {\it HST} $I$-band data of Lauer {\it et al.} (1992).
We use their seeing-deconvolved profile,
and approximate the photometric uncertainties by
$\Delta \mu = {\rm max}[0\parcs011 /R,0.02]$.
For the outer regions ($R$=8\arcs--745\arcs),
we use the $B$-band data of Capaccioli, Caon, \& Rampazzo (1990),
with the photometric uncertainties estimated from their Figure 3, and
with an offset of 2.78 magnitudes
to match the profile to the Lauer {\it et al.} (1992) data.
The color gradients are quite small over the entire range of the galaxy
($|\Delta (B-I)|_{\rm max} < 0.2$ mag;
{\it e.g.}, Boroson {\it et al.} 1983; Cohen 1986; 
Zeilinger {\it et al.} 1993),
so there should be little systematic problem in combining the $I$-band and $B$-band data.
The isophote ellipticity varies from 
0.02 at $2\arcs$
to 0.1 at $75\arcs$
to 0.35 at $500\arcs$.
The dynamics of the galaxy depend on the ellipticity of its gravitational
potential, which is much rounder than its density distribution
($\epsilon_\Phi \sim \epsilon_\rho/3$; Binney \& Tremaine 1987 \S 2.3.1),
so it will be a reasonable approximation to treat the galaxy as spherical.
We map all of the data from the major axis $a$ 
to the intermediate axis $m=a\sqrt{1-\epsilon}$.

For the purposes of our mass modeling, we project and fit 
a parameterized luminosity density model to the data of the form
$\nu_\star(r) = \nu_0 (r/s_1)^{-\alpha_1}(1+r^3/s_2^3)^{-\alpha_2}(1+r/s_3)^{-\alpha_3}$,
and find 
$\nu_0=9.9\times10^6 L_{B,\odot}$~arcsec$^{-3}$,
$\alpha_1 = 1.22$, $\alpha_2 = 0.361$, $\alpha_3=1.26$, 
$s_1=1\arcs$, $s_2=9\parcs38$, and $s_3=170\arcs$,
with a $\chi^2$ statistic for the fit of 35.1 for 142 degrees of freedom.
This functional fit will only be used to generate a constant $M/L$ gravitational potential,
while we will directly fit the surface brightness data in our dynamical models (\S 3).
Note that in M87, $\mu(R)$ does not taper off steeply in the outer regions
as in normal $R^{1/4}$ ellipticals ---
a typical $R^{-2}$ cD envelope (Schombert 1986) begins at $R \gsim 300\arcs$.
An old dispute over the brightness of this envelope
(see {\it e.g.}, de Vaucouleurs \& Nieto 1978; Carter \& Dixon 1978)
has not been resolved,
presumably because of sky-subtraction problems for such a
large, low-surface-brightness structure.
Our asymptotic radial slope lies
between the reported extremes.

\subsection{Stellar Velocities}

For the core region ($R < 29\arcs$), 
we use the {\it G}-band data from van der Marel (1994, hereafter vdM)
for the projected stellar velocity dispersion $\hat{\sigma}_{\rm p}(R)$ and the
Gauss-Hermite velocity moments $h_4(R)$ and $h_6(R)$
(see \S 3).
For the outer regions ($R$=28\arcs--168\arcs),
we use $\hat{\sigma}_{\rm p}(R)$ from Sembach \& Tonry (1996, hereafter ST).
We combine the data from positive and negative radii into one radial profile.
There appears to be a systematic velocity offset between the ST data and most other data
(Sargent {\it et al.} 1978; Davies \& Birkinshaw 1988; 
Jarvis \& Peletier 1991; Winsall \& Freeman 1993),
presumably due to the large slit width;
we estimate that this corresponds to an additional instrumental dispersion
of $183\pm11$ km s$^{-1}$,
which we remove from the ST data in order to match them to the vdM data.
The final velocity dispersion profile is nearly constant for $R<1\arcs$,
outside of which it falls off slowly with radius
($\hat{\sigma}_p \sim R^{-0.1}$).
Comparing $h_4$ and $h_6$ from positive and negative radii,
we find differences larger than are consistent with the stated uncertainties.
We assume the errors were underestimated by 12\% and 22\% to bring the profiles into
statistical agreement.
The departures from Gaussianity of the LOSVD are everywhere small
(typically, $|h_4|, |h_6| < 0.02$).

\subsection{Globular Cluster Surface Density}
Our GC surface density radial profile $N(R)$ is taken from the number 
counts of Kundu {\it et al.} (1999) for $R=0\arcs$--$96\arcs$
(with their quoted uncertainties).
For $R=84\arcs$--$472\arcs$,
we take the data from McLaughlin, Harris, \& Hanes (1993),
adding in their 0.4 arcmin$^{-2}$ background uncertainty,
and multiplying by 1.45 to match the normalization of the Kundu {\it et al.} (1999) data.
For $R=419\arcs$--$1351\arcs$,
we use the data from Harris (1986)
after subtracting their background count of $5.8\pm0.3$ arcmin$^{-2}$
and normalizing by a factor of 2.19.
We do not convert any of these profiles to an intermediate axis since this is already
effectively accomplished by the derivation method (counting in circular annuli).
Like the stellar surface brightness, the GC surface density
decreases slowly with radius in the central regions ($\sim R^{-0.3}$), 
and changes over to a steeper power law in the outer parts ($\sim R^{-1.6}$);
however, the radius where this break occurs is much larger for the GCs
($\sim 60\arcs$ vs. $\sim 10\arcs$),
demonstrating that these systems are dynamically distinct.

\subsection{Globular Cluster Discrete Velocities}

For the GC line-of-sight velocities $v_z$,
we use data from
Mould, Oke, \& Nemec (1987), Huchra \& Brodie (1987), Mould {\it et al.} (1990),
CR, and Cohen (2000).
Most of the positions are taken from Strom {\it et al.} (1981).
We compare the velocities of common objects to determine the
systematic offset between the data sets,
and to estimate the measurement uncertainties 
($\Delta v_z \sim 100$ km s$^{-1}$).
We discard as foreground stars all objects with 
heliocentric velocities $v_z < 250$ km s$^{-1}$,
and as background galaxies all those with $v_z > 2650$ km s$^{-1}$.
We examine the colors and magnitudes of
objects with $v_z < 500$ km s$^{-1}$ to distinguish GCs from stars.
Objects of uncertain identification are discarded.
We map the data to the intermediate axis $m$,
using ellipticity estimates from McLaughlin, Harris, \& Hanes (1994).
Our final data set has 234 velocities from $R=25\arcs$--$526\arcs$
(see Fig. 1).

{\plotfiddle{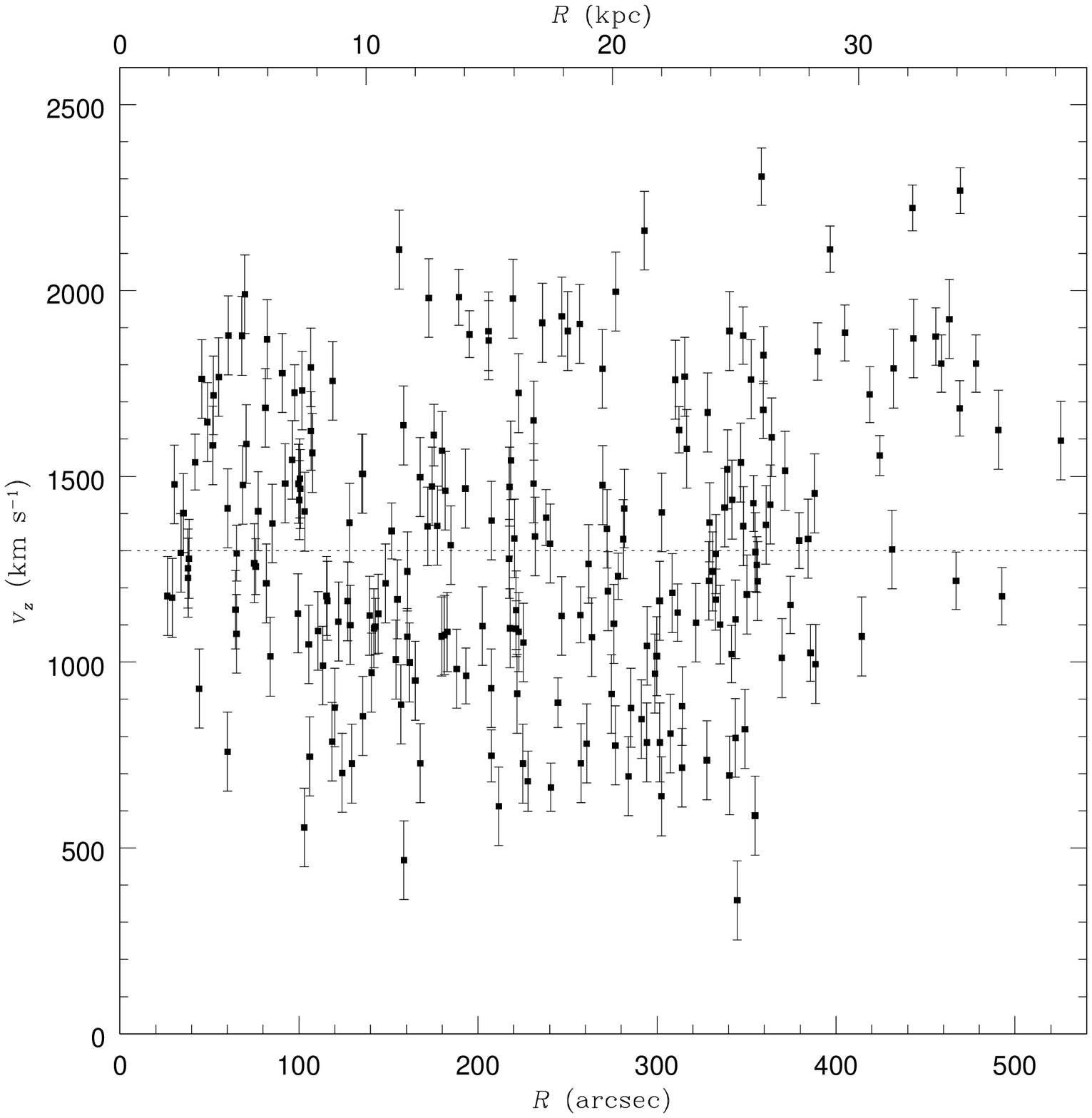}{5.0in}{0}{68}{68}{18}{-120}}

{\center \small {\bf F{\scriptsize \bf IG}. 1.---}
Line-of-sight velocities for 234 globular clusters around M87, 
as a function of galactocentric radius $R$.
The dotted line shows a systemic velocity of 1300 km s$^{-1}$.
}

There is a
paucity of low velocity measurements at large radii,
which may be related to
the rotation of the GCs along the galaxy's major axis in the outer parts,
as demonstrated by Kissler-Patig \& Gebhardt (1998).
For the GCs with $R\geq 370\arcs$,
there were only 4 measurements taken
on the ``approaching'' side
and 23 on the ``receding'' side.
Thus, the large-radius velocity asymmetry is probably due to incomplete spatial coverage,
which may make our mean velocity for the system systematically high
(it is decreased from 1352 to 1293 km s$^{-1}$ if the clusters with
$R \geq 370\arcs$ are omitted).
This is also suggested by the global velocity distribution (Fig. 2),
which appears more symmetric if the mean velocity is 
set to $\sim 1300$ km s$^{-1}$.
The best-fit Gaussian curve to the data has a central velocity of
$\hat{v}_{\rm p} = 1323\pm28$ km s$^{-1}$ ($1308\pm29$ km s$^{-1}$ if the outer clusters are
omitted).
We adopt a systemic velocity of 1300 km s$^{-1}$,
which we subtract from the measurements to get the galactocentric velocities.
This compares
well with estimates of the stellar mean redshift from vdM (1277 km s$^{-1}$),
ST (1293 km s$^{-1}$), and the RC3 ($1282\pm9$ km s$^{-1}$; de Vaucouleurs {\it et al.} 1991).
The large-radius velocity asymmetry will not pose any further difficulties
because we will model only the {\it even} part of the DF,
so that all the discrete velocities are implicitly 
considered as present also at their reflected position
about the mean velocity.

Fig. 2 also suggests that the GCs have a tangentially-biased orbit
distribution,
which typically produces a flat or double-peaked LOSVD.
We fit a fourth-order Gauss-Hermite velocity moment (see \S 3) to
the system, finding $h_4=-0.04\pm0.02$ 
($\hat{\sigma}_{\rm p}=404\pm16$ km s$^{-1}$),
which is weakly suggestive of tangential anisotropy.
A double-peaked LOSVD (especially evident at larger radii)
has also been found in the GC system of the Fornax Cluster cD galaxy NGC 1399 
(Kissler-Patig {\it et al.} 1999).

{\plotfiddle{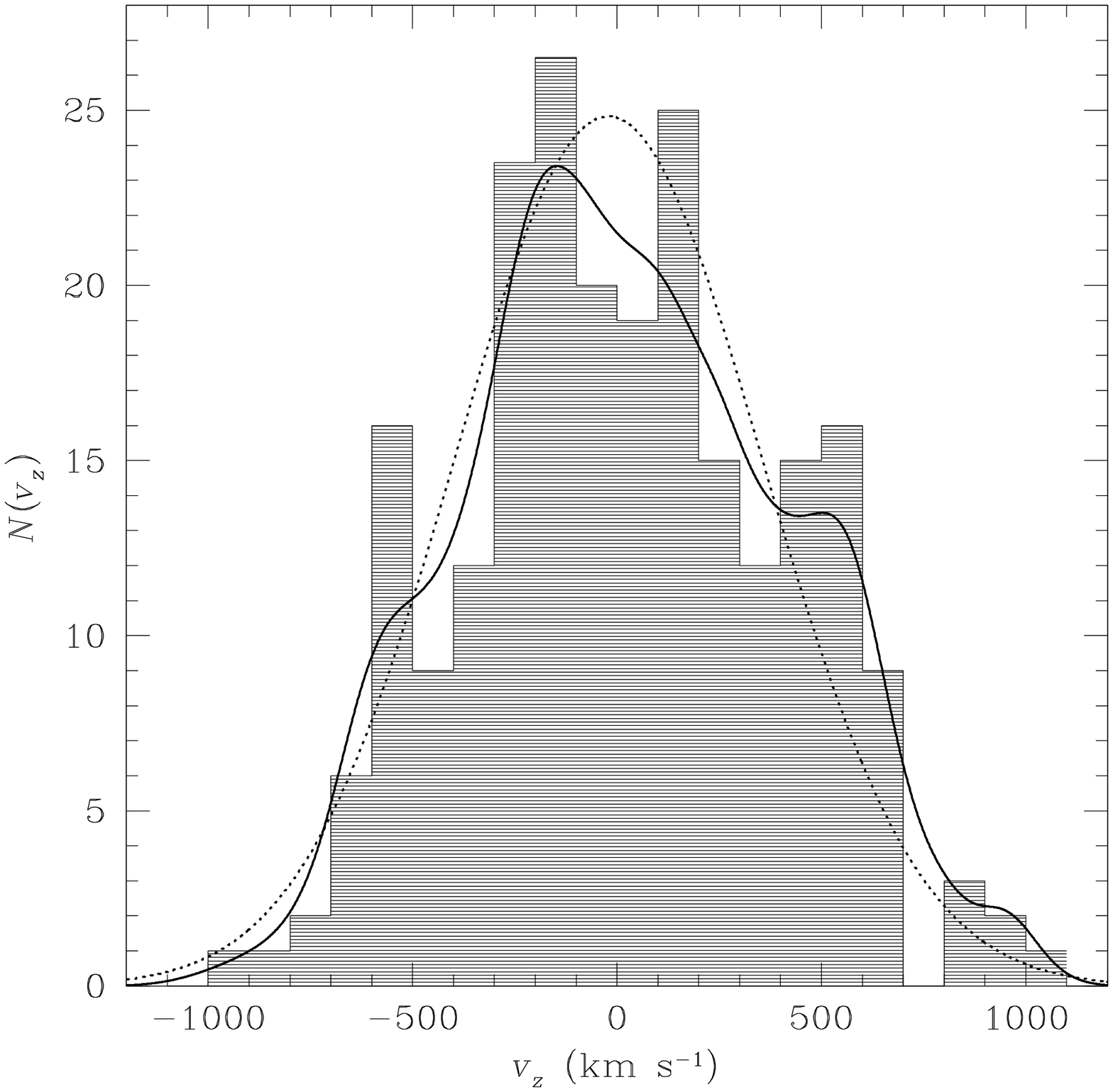}{2.64in}{0}{40}{40}{100}{-77}}

{\center \small {\bf F{\scriptsize \bf IG}. 2.---}
Distribution of globular cluster line-of-sight velocities,
relative to a systemic velocity of 1300 km s$^{-1}$.
The histogram shows the number of data points in velocity bins.
The solid curve shows a smoothed superposition of the measurements.
The dotted curve shows the best-fit Gaussian curve to the data
($\sigma=376$ km s$^{-1}$).
Note the
double peaks of low significance
at $\pm 200$ km s$^{-1}$ and at $\pm 500$ km s$^{-1}$.
}

\section{METHODS}

The method of orbit modeling was pioneered by
Schwarzschild (1979) to reproduce the observed density distribution of a galaxy
from a weighted library of representative particle orbits.
Given a fixed gravitational potential $\Phi(r)$,
the solutions are nonparametric ---
no functional form is assumed for the DF ---
so that one need not worry that the models cannot fully explore the 
possible solutions.
The DF is guaranteed to be nonnegative and physical.
This method was extended by Richstone \& Tremaine (1984) to include
projected velocity dispersion information,
and by Rix {\it et al.} (1997) to include higher-order velocity moments.
We used similar techniques
to examine the uncertainty in a
high-redshift galaxy's mass as implied by a measurement of its central velocity
dispersion --- an important step in determining $H_0$ from
certain gravitational lens systems (Romanowsky \& Kochanek 1999, hereafter Paper I).
Other recent orbit modeling efforts have concentrated on applying the method
to three-integral axisymmetric systems 
({\it e.g.}, van der Marel {\it et al.} 1998; 
Cretton \& van den Bosch 1999;
Gebhardt {\it et al.} 2000;
Cretton, Rix, \& de Zeeuw 2000).
Here we extend our spherical method to include the discrete
velocities of GCs.
The outline of the method is as follows.

We begin with an assumed radial profile for the mass density distribution $\rho(r)$.
Our first model is a simple representation of a dark-matter dominated system,
the singular isothermal sphere (SIS):
$\rho(r) = \sigma_0^2 / 2 \pi G r^2$.
The second model is the constant $M/L$ distribution described in \S 2.1,
parameterized by $\Upsilon_B$:
the ratio of the mass density $\rho(r)$ and the $B$-band luminosity density $\nu_\star(r)$,
with units of $M_\odot/L_{\odot,B}$.
We pick a random distribution of orbits that densely samples
the phase space of energy and angular momentum $(E,|L|)$.
The initial radii $r_{0k}$ of the orbits are logarithmically spaced in $r_0$,
and the energy $E_k$ of each orbit is selected to correspond to that of a
circular orbit at this radius, $\Phi(r_{0k})+v_{\rm c}^2(r_{0k})/2$.
For the SIS model, this procedure produces a sampling that is uniform in energy.
The angular momentum $L_k$ of the orbit is selected randomly from the
range $[0,L_{{\rm max},k}]$, where $L_{{\rm max},k}=r_{0k} v_{\rm c}(r_{0k})$.
For both the stellar models and the GC models,
we use 2500 particle orbits.
For the stars, $r_0$ spans the range $0\parcs07$--$4450\arcs$,
resulting in a radial coverage of $r=0\arcs$--$7330\arcs$ (SIS model)
or $r=0\arcs$--$8710\arcs$ (constant $M/L$ model).
For the GCs, $r_0=$10\parcs0--$13360\arcs$,
corresponding to $r=0\arcs$--$22030\arcs$ (SIS model)
or $r=0\arcs$--$26420\arcs$ (constant $M/L$ model).

We next compute the orbit projection ``kernels'',
which correspond to the contribution of each orbit to each observable.
The predictions of the model can then be expressed as follows:
\begin{equation}
y_i^{\rm m} = \sum_k w_k^2 \langle K_{ik} \rangle_{t,\theta,\phi} ,
\end{equation}
where the ${\bf w}$ are the orbit weights, and the kernels ${\bf K}$ have
been averaged over time and over all spherical-polar viewing angles
$(\theta,\phi)$.
For example, the kernel for the angle-averaged surface density in an annulus
between $R_1$ and $R_2$
of an orbit at an instantaneous radius $r^\prime$ is given by the integral
\begin{eqnarray}
\langle K_{I(R_1,R_2)} \rangle_{\theta,\phi} (r^\prime) &=& \frac{1}{\pi (R_2^2-R_1^2)} \int_{R_1}^{R_2} 2\pi R dR \int_{R}^{\infty} \frac{\delta(r-r^\prime)}{4\pi r^2} \frac{2 r dr}{\sqrt{r^2-R^2}} \int_0^{2\pi} \frac{d\xi}{2\pi} \\
&=& \frac{\sqrt{1-R_1^2/r^{\prime 2}}-\sqrt{1-R_2^2/r^{\prime 2}}}{\pi (R_2^2-R_1^2)} ,
\end{eqnarray}
where the orbit's instantaneous density is $\delta(r-r^\prime)/4\pi r^2$,
the variable $\xi$ represents
the line of sight's position in the tangential plane $(\theta,\phi)$,
and the integration is carried out along the line of sight $z = \pm \sqrt{r^2-R^2}$.
The viewing-angle integral (along $d\xi$) is nontrivial for kernels that
involve velocity measurements.
Placed at the initial radius $r_0$, 
the orbit is run forward in time for one radial period $T_r$, and the final
kernel is found by averaging over time:
\[
\langle K \rangle_{t,\theta,\phi} = \frac{1}{T_r} \int_0^{T_r} \langle K \rangle_{\theta,\phi}(t) dt ,
\]
where the integration is handled by 
a Bulirsch-Stoer type integrator.

We must calculate the model's LOSVD,
$dL/dv_{\rm p}(v_{\rm p},R)$,
to fit the observed stellar
velocity moments ($\hat{\sigma}_{\rm p},h_4,h_6$) and the GC 
velocity measurements.
The LOSVD kernels $K_{dL/dv_{\rm p}(R), k}$
for a measurement at radius $R$
are binned in velocity with $v_{\rm p}=[0..v_{\rm max}]$,
where $v_{\rm max}(R)$ is the velocity of a particle at radius $r=R$ that has
plunged inward on a radial orbit from the largest possible radius.
For the constant $M/L$ model, this implies that $v_{\rm max} \simeq v_{\rm esc}$
(the escape velocity).
Our models are symmetric in velocity,
{\it i.e.}, $dL/dv_{\rm p}(v_{\rm p})=dL/dv_{\rm p}(-v_{\rm p})$.
By construction, $v^2_{\rm max}$ scales linearly with the mass normalization of the
galaxy model ($\sigma^2_0$ or $\Upsilon_B$), so that we can use the same
binned LOSVD kernels when changing the mass scale.

The Gauss-Hermite velocity moments $h_l$, which measure deviations
of a LOSVD from a Gaussian, are defined by
\begin{equation}
h_l \equiv \frac{\sqrt{2}\gamma_0}{\hat{\gamma}_{\rm p}}\int_{-\infty}^\infty \frac{dL}{dv_{\rm p}}(v_{\rm p}) e^{-\hat{w}^2/2} H_l(\hat{w}) dv_{\rm p} ,
\end{equation}
where $\hat{w}=(v_{\rm p}-\hat{v}_{\rm p})/\hat{\sigma}_{\rm p}$,
$\gamma_0$ is the line strength,
$(\hat{\gamma}_{\rm p},\hat{v}_{\rm p},\hat{\sigma}_{\rm p})$ are the
coefficients for the best Gaussian fit to $dL/dv_{\rm p}$,
and $H_l(\hat{w})$ are the Hermite polynomials
(van der Marel \& Franx 1993).
The moments ($h_4$, $h_6$) measure how flat ($h_4<0$, $h_6>0$)
or how peaked ($h_4>0, h_6<0$) the LOSVD is compared to a Gaussian,
where these signatures are typically produced by tangential
and radial orbits, respectively.
The two moments differ in how much weight they put in the high-velocity
wings of the LOSVD.
In Paper I,
we fitted a Gaussian curve to the LOSVD to find 
($\hat{\gamma}$, $\hat{\sigma}_{\rm p}$)
at each step in our minimization routine,
and then calculated $h_4$;
this is equivalent to solving the nonlinear equations $h_0=1$ and $h_2=0$.
Here, we expand the Gauss-Hermite series about the fixed data values
$\hat{\sigma}^{\rm d}_{\rm p}$.
Rather than directly fitting $\hat{\sigma}^{\rm d}_{\rm p}$,
we equivalently fit the second-order moment
$h_2 = 0.00\pm\Delta h_2$, where
$\Delta h_2 \simeq \Delta \hat{\sigma}_{\rm p} / \sqrt{2}\hat{\sigma}^{\rm d}_{\rm p}$ .
With this formulation,
calculating the model's velocity observables is now linear with respect
to the orbit weights.
We use 41 velocity bins from $v_{\rm p}=0$ to $v_{\rm max}$ 
to numerically integrate equation (4),
which we have heuristically found to give very good accuracy.
Note that because our model is completely spherically symmetric
(simulating only the even part of the DF),
it would be more appropriate to fit measurements of
$z_4$ (expanded around $\hat{v}_{\rm p}=0$) rather than $h_4$
(see van der Marel {\it et al.} 1994 \S 5.1);
however, these measurements are not available for M87.
Since the measured rotation is small ($\sim 10$ km s$^{-1}$),
the difference would be unimportant.

The kernels $({\bf K}_{I}, {\bf K}_{dL/dv_{\rm p}})$
need be computed only once for each galaxy mass model.
We then adjust the weights of the orbits so that the model's
projected observables ${\bf y}^{\rm m}$ (equation 1) best fit the data ${\bf y}^{\rm d}$.
For most of the data ($I,\hat{\sigma}_{\rm p},h_4,h_6$),
we express the likelihood of the fit using the standard $\chi^2$ statistic,
\begin{equation}
\chi^2 = \sum_i \left( \frac{y^{\rm m}_i - y^{\rm d}_i}{\sigma_i}\right)^2 ,
\end{equation}
where the measurement uncertainty is $\sigma_i$,
while the likelihood function for the discrete velocities and positions
of the GCs is
\begin{equation}
{\cal L}_i(v_i,R_i) \propto \int \frac{dL}{dv_{\rm p}} (v_{\rm p},R_i) e^{-(v_i-v_{\rm p})^2/2\sigma_i^2} dv_{\rm p} ,
\end{equation}
where $v_i \pm \sigma_i$ are the individual velocity measurements.
In maximizing ${\cal L}_i$, we are forcing the LOSVD to peak at the
measured velocities $v_i$ weighted by the measurement errors $\sigma_i$.
This is schematically illustrated in Fig. 3.
Given complete freedom, this method would produce a best-fit solution whose
LOSVDs resembled $\delta$-functions at the measurements $v_i$.
However, the model's averaging in angle and time produces an intrinsic smoothing
to the LOSVDs, and does not permit such unphysical solutions.
For the current implementation, we use 15 velocity bins
from $v_{\rm p}=0$ to $v_{\rm max}$
for the LOSVD
at each radius in order to compute the integral in equation (6).
Note that our method does not require binning the velocities
in radius, nor computing velocity moments.

{\plotfiddle{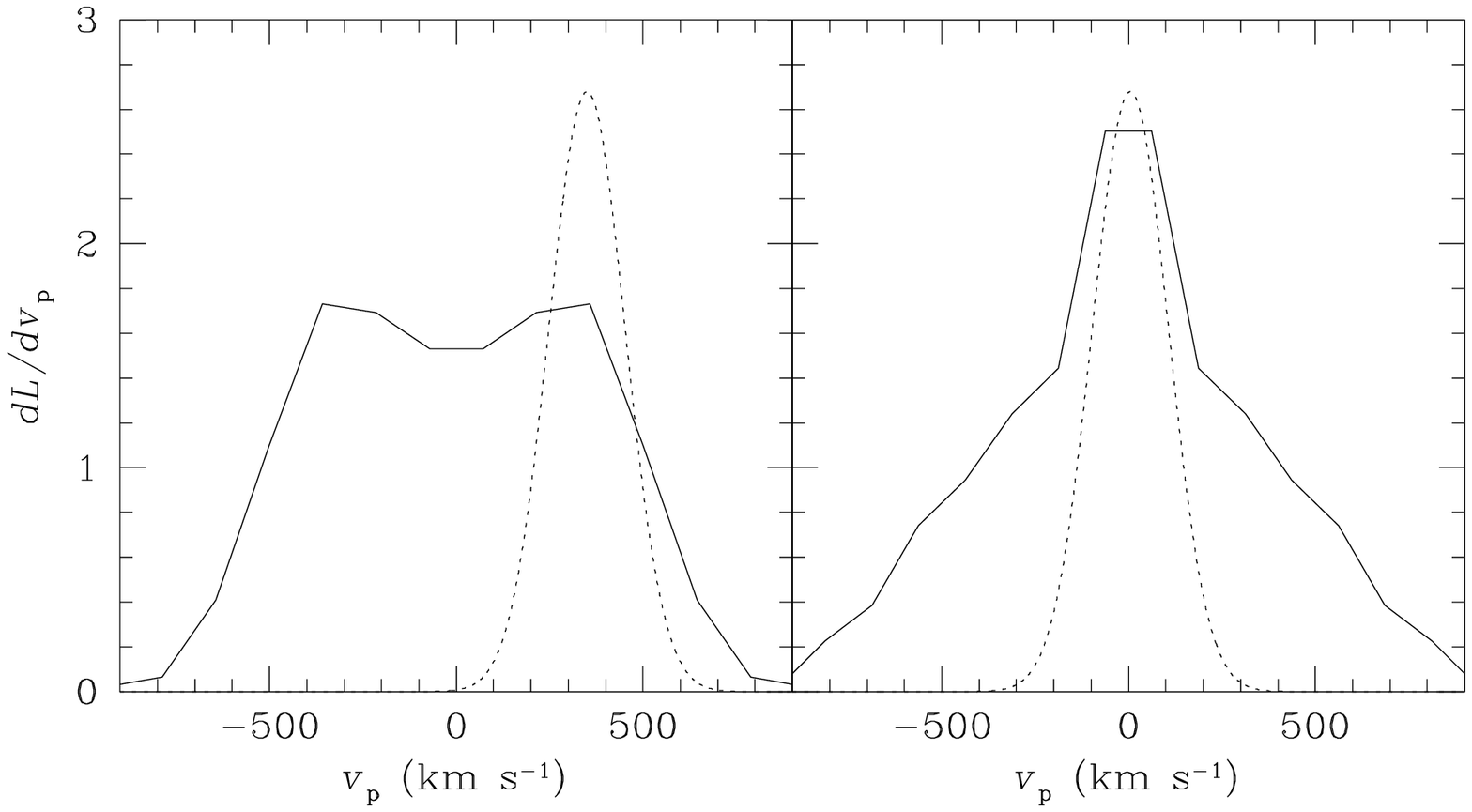}{1.9in}{0}{57}{57}{52}{-112}}

{\center \small {\bf F{\scriptsize \bf IG}. 3.---}
Schematic line-of-sight velocity distributions ({\it solid curves}),
fitted to discrete velocity data ({\it dotted curves}).
The double-peaked LOSVD on the left is indicative of tangential anisotropy,
while the centrally-peaked LOSVD on the right indicates radial anisotropy.
}

\vspace{10pt}

The final function which we will minimize is
\begin{equation}
f({\bf w}) \equiv \frac{1}{2} \chi^2 - \sum_i \ln{\cal L}_i + \lambda S ,
\end{equation}
where $S$ is a measure of entropy:
\begin{equation}
S = \sum_k w_k^2 \ln w^2_k .
\end{equation}
We employ this smoothing function $S$ as a heuristic device for quickly
reaching a rough solution; in our final solutions, the Lagrangian factor
$\lambda$ is reduced to such a low value $(\sim 10^{-5})$ so as to make the
entropy constraints inconsequential.
With no such regularization imposed, this is an ill-conditioned inversion problem,
so our DF solutions will be choppy in a way that real galaxies' DFs
presumably are not.
However, our methods are the statistically correct way to handle the
uncertainties, absent any other {\it a priori} smoothness conditions.
We use a conjugate gradient method (see Press {\it et al.} 1992 \S 10.6)
with first and second derivative information, to minimize $f$.
We have tested our methods on an isotropic Hernquist (1990) model to demonstrate that we
recover correctly the input mass, given constraints on $I(R)$ and $\hat{\sigma}_{\rm p}(R)$
(see Paper I).

\section{RESULTS}

Here we present the results of our modeling.
First we investigate the necessity for dark matter in the system
by comparing two simple models
(\S 4.1).
Then we consider more generalized mass models (\S 4.2).
Finally, we report constraints on the orbit anisotropies (\S 4.3).

\subsection{Test for Dark Matter}

We first fit the two simple mass models (SIS and constant $M/L$) to the stellar data
[$I(R)$, $\hat{\sigma}_{\rm p}(R)$, $h_4(R)$, $h_6(R)$] over a sequence of
the mass parameters $\sigma_0$ and $\Upsilon_B$
(see Fig. 4).
Table 1 lists the best-fit values for these parameters
and the difference in log-likelihood between the two models.
For the SIS model, we find
$\sigma_0 = 272^{+19}_{-5} $ km s$^{-1}$;
the fit of the best solution to the data is shown in Fig. 5.
For the constant $M/L$ model, we find
$\Upsilon_B = 8.1\pm0.6$.
This model does not fit the data as well as the best SIS model does
(1 $\sigma$ significance),
providing weak evidence for a dark matter halo in this galaxy.

In order to see which constraints are important for indicating the 
presence of dark matter, we have tried models which lack either
the $h_6(R)$ data 
or the large-radius $\hat{\sigma}_{\rm p}(R)$ data,
(Table 1).
Dropping these constraints makes little difference in
differentiating between mass models
(though the mass normalization of the SIS model does change in the latter case).
In particular, we note that the extra velocity dispersion data is unhelpful,
owing to the flexibility of the orbital anisotropies to reproduce
an arbitrary velocity dispersion profile.
Only with additional LOSVD information at large radii can the stellar
kinematical data by itself further constrain the mass distribution.

{\plotfiddle{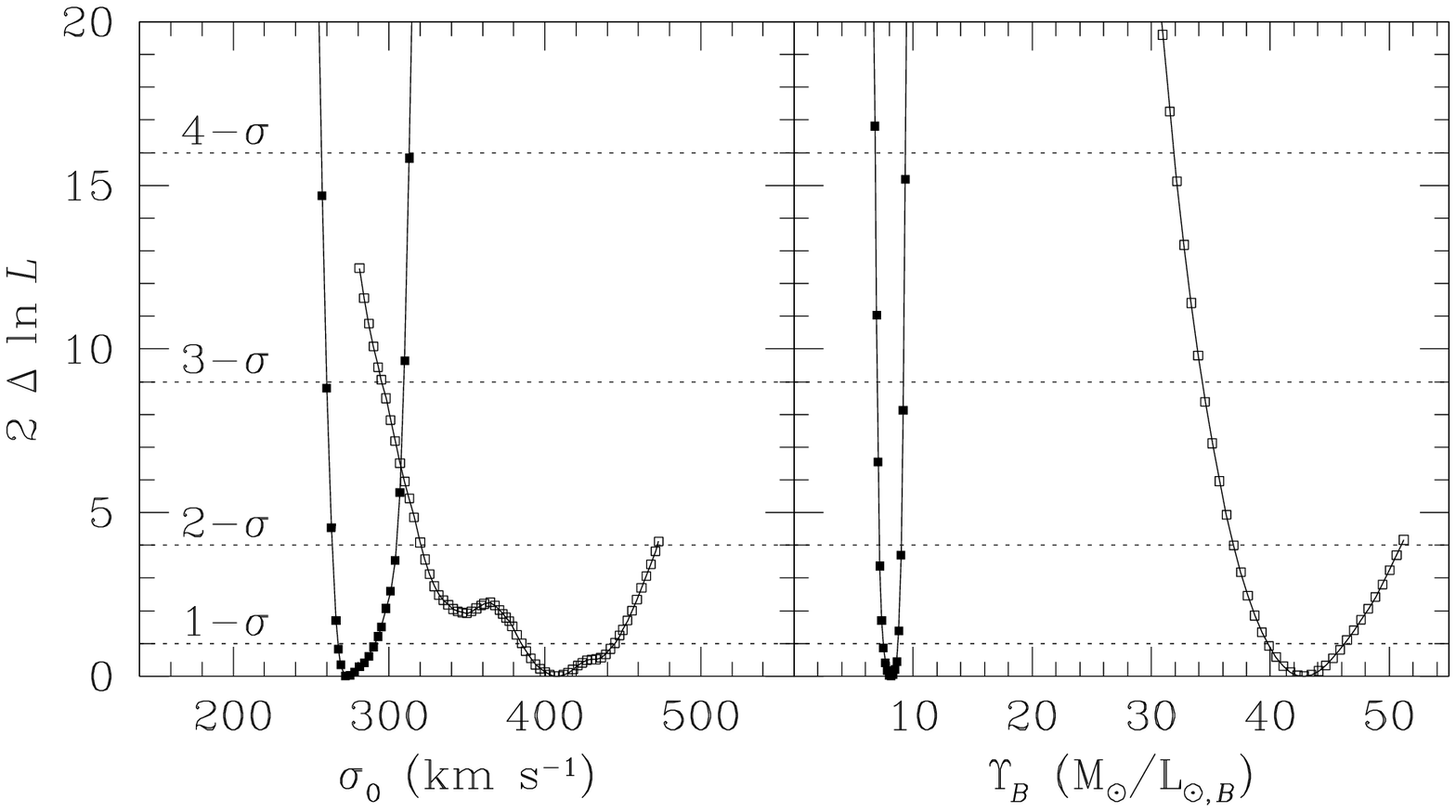}{2.85in}{0}{80}{80}{-20}{-332}}

{\center \small {\bf F{\scriptsize \bf IG}. 4.---}
The log-likelihood of
the best-fit singular isothermal model ({\it left panel})
and constant mass-to-light ratio model ({\it right panel}),
as a function of the mass parameters $\sigma_0$ and $\Upsilon_B$.
The filled squares show the fit to the stellar data, and the open squares the fit to
the globular cluster data.
}

\begin{tabular}{lcrrr}
\\
\hline\hline\\
& Singular Isothermal &   &
\multicolumn{2}{c}{Constant $M/L$} \\
\cline{2-2} \cline{4-5} \\
Constraints & $\sigma_0$ (km s$^{-1}$) & 
 & \multicolumn{1}{c}{$\Upsilon_B$ ($M_\odot/L_{\odot,B}$)} & \multicolumn{1}{c}{$\Delta \ln {\cal L}$} \\ \\
\hline \\
\vspace{4pt}
Stars: $I$, $\hat{\sigma}_{\rm vdM}$, $\hat{\sigma}_{\rm ST}$, $h_4$, $h_6$ & $272^{+19}_{-5}$ & 
& $8.1\pm0.6$ & 0.8 (1.0 $\sigma$) \\
\vspace{4pt}
Stars: $I$, $\hat{\sigma}_{\rm vdM}$, $\hat{\sigma}_{\rm ST}$, $h_4$ & $275^{+19}_{-9} $ & 
& $8.1^{+0.7}_{-0.6}$ & 0.6 (0.9 $\sigma$) \\
\vspace{4pt}
Stars: $I$, $\hat{\sigma}_{\rm vdM}$, $h_4$, $h_6$ & $254^{+22}_{-18}$ & 
& $8.3^{+1.0}_{-1.7}$ & 0.9 (1.0 $\sigma$) \\
\vspace{4pt}
Globular clusters & $409^{+36}_{-24}$ &
& $42.9^{+3.3}_{-3.1}$ & 12.9 (4.7 $\sigma$) \\ 
\vspace{4pt}
Stars + Globular clusters & $301^{+5}_{-14}$ & 
& $9.3^{+0.1}_{-0.2}$ & 206.0 (20.1 $\sigma$) \\  \\
\hline
\end{tabular}
{\center \small {\bf T{\scriptsize \bf ABLE} 1.---}
M87 model solutions, given different sets of observational constraints.
The best-fit mass parameters ($\sigma_0$ and $\Upsilon_B$) are given for the
SIS and constant $M/L$ models.
The difference in log-likelihood between the two models is shown,
along with the statistical significance at which the SIS model is preferred.
The ``Bayesian'' probability for the SIS model is $(1+e^{-\Delta \ln {\cal L}})^{-1}\simeq 1$,
while that for the constant $M/L$ model is $e^{-\Delta \ln {\cal L}}(1+e^{-\Delta \ln {\cal L}})^{-1} \simeq e^{-\Delta \ln {\cal L}}$
for $\Delta \ln {\cal L} \gsim 2$.
}

{\plotfiddle{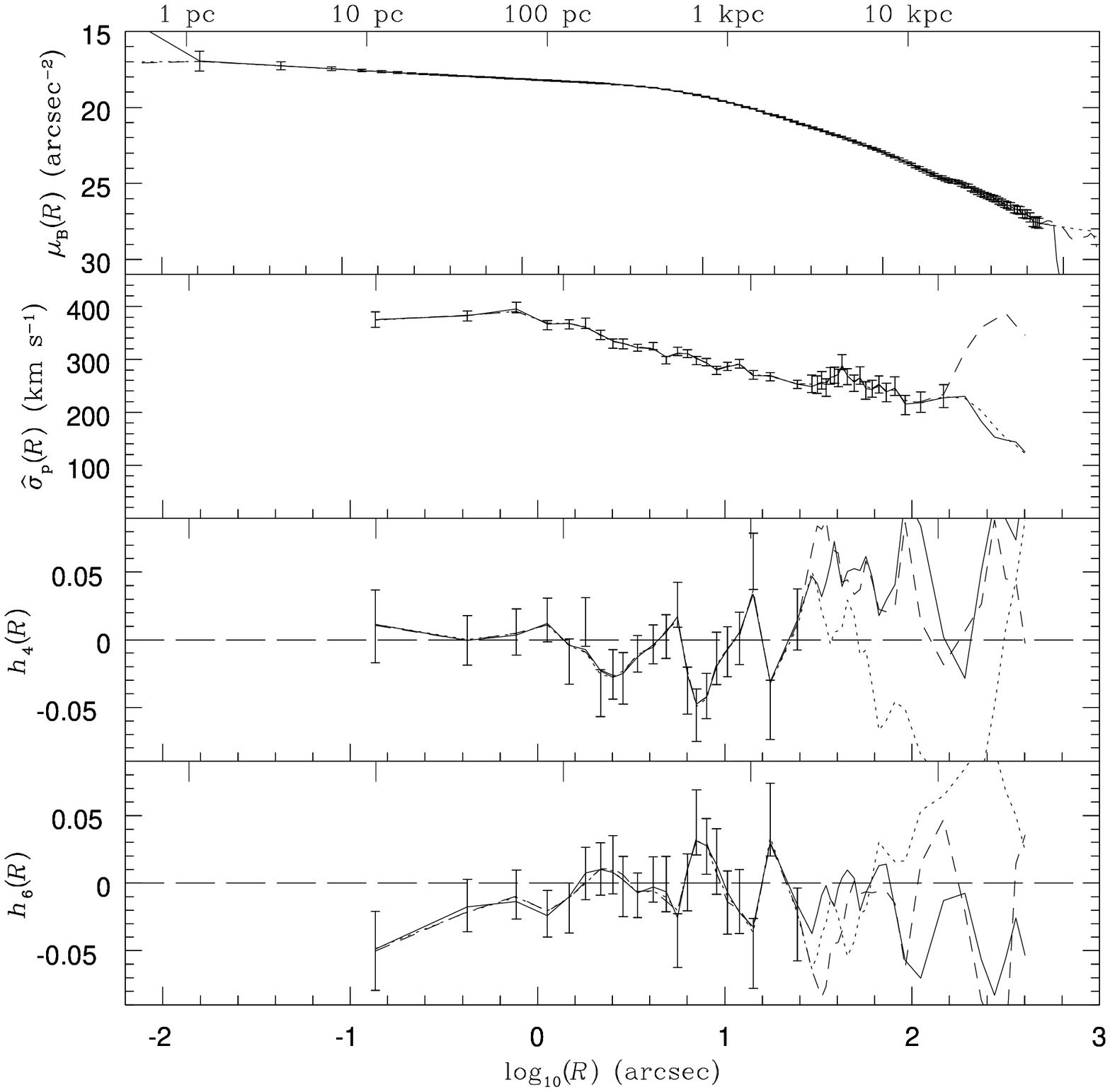}{7.0in}{0}{90}{90}{-35}{-145}}

{\center \small {\bf F{\scriptsize \bf IG}. 5.---}
Fits to the M87 stellar data ({\it error bars})
for the best singular isothermal model 
($\sigma_0=272$ km s$^{-1}$ : {\it solid lines}),
the best constant mass-to-light ratio model 
($\Upsilon_B=8.1$ : {\it dotted lines}),
and the best NFW2 model (see \S 4.2; {\it dashed lines}).
The data are the radial profiles of 
the surface brightness, the velocity dispersion,
and the fourth- and sixth-order Gauss-Hermite velocity moments
({\it from top to bottom}).
Note that $R_{\rm eff}\simeq 100\arcs$.
}

We fit the GC data to the same two mass models (see Fig. 4 and Table 1).
For the SIS model, 
we find $\sigma_0=409^{+36}_{-24}$ km s$^{-1}$;
the fit of the best SIS solution to the data is shown in Figs. 6--8.
Note especially in Fig. 7 that some of the higher-order features of the observed LOSVDs are
reproduced by the model.
For the constant $M/L$ model, we find $\Upsilon_B = 43\pm3$.
This model fits the data much worse than does the SIS model
(5~$\sigma$ significance).
Here, the constant $M/L$ model cannot reproduce the radially-rising GC velocity dispersion (see Fig. 6).
Also, $\Upsilon_B=43$
is implausible for a standard stellar population,
requiring an age $\gg 17$ Gyr and a metallicity $[{\rm Fe/H}] \gg 0.50$ (Worthey 1994),
so that even if mass traced light, the mass could not 
consist of a standard stellar population.
Finally, note that $\Upsilon_B=43\pm3$ is completely inconsistent with $8.1\pm0.6$
as obtained via the stellar dynamics, unless there is a drastic radial gradient in
the stellar population.
We thus conclude that the constant $M/L$ model is definitively ruled out for M87,
assuming standard gravitational dynamics.
(Note that the 20~$\sigma$ significance given in Table 1 is formal only,
and that in practice, a number of systematic uncertainties will reduce this
significance ---
but it can still be said that the result holds to a high significance level.)

{\plotfiddle{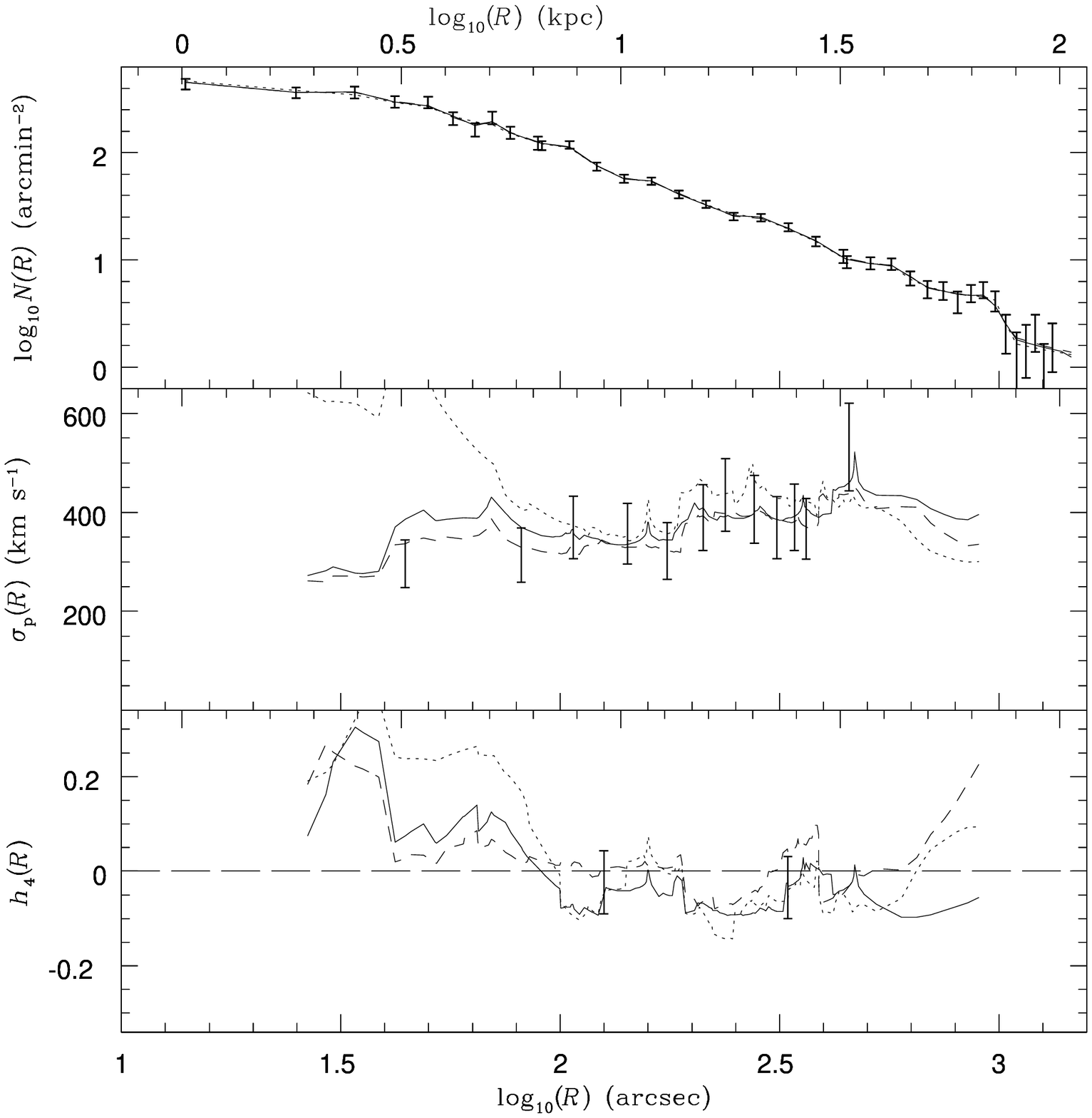}{6.8in}{0}{88}{88}{-35}{-145}}

{\center \small {\bf F{\scriptsize \bf IG}. 6.---}
Fits to the M87 globular cluster data ({\it error bars})
for the best singular isothermal model ( $\sigma_0=409$ km s$^{-1}$: {\it solid lines}),
the best constant mass-to-light ratio model ($\Upsilon_B = 42.9$: {\it dotted lines}),
and the best NFW2 model (see \S 4.2; {\it dashed lines}).
The data include the radial profiles of 
the surface density, the velocity dispersion,
and the fourth-order Gauss-Hermite velocity moment
({\it from top to bottom}).
Note that only the surface density is actually fit, while the other
data are shown for comparative purposes only.
}

{\plotfiddle{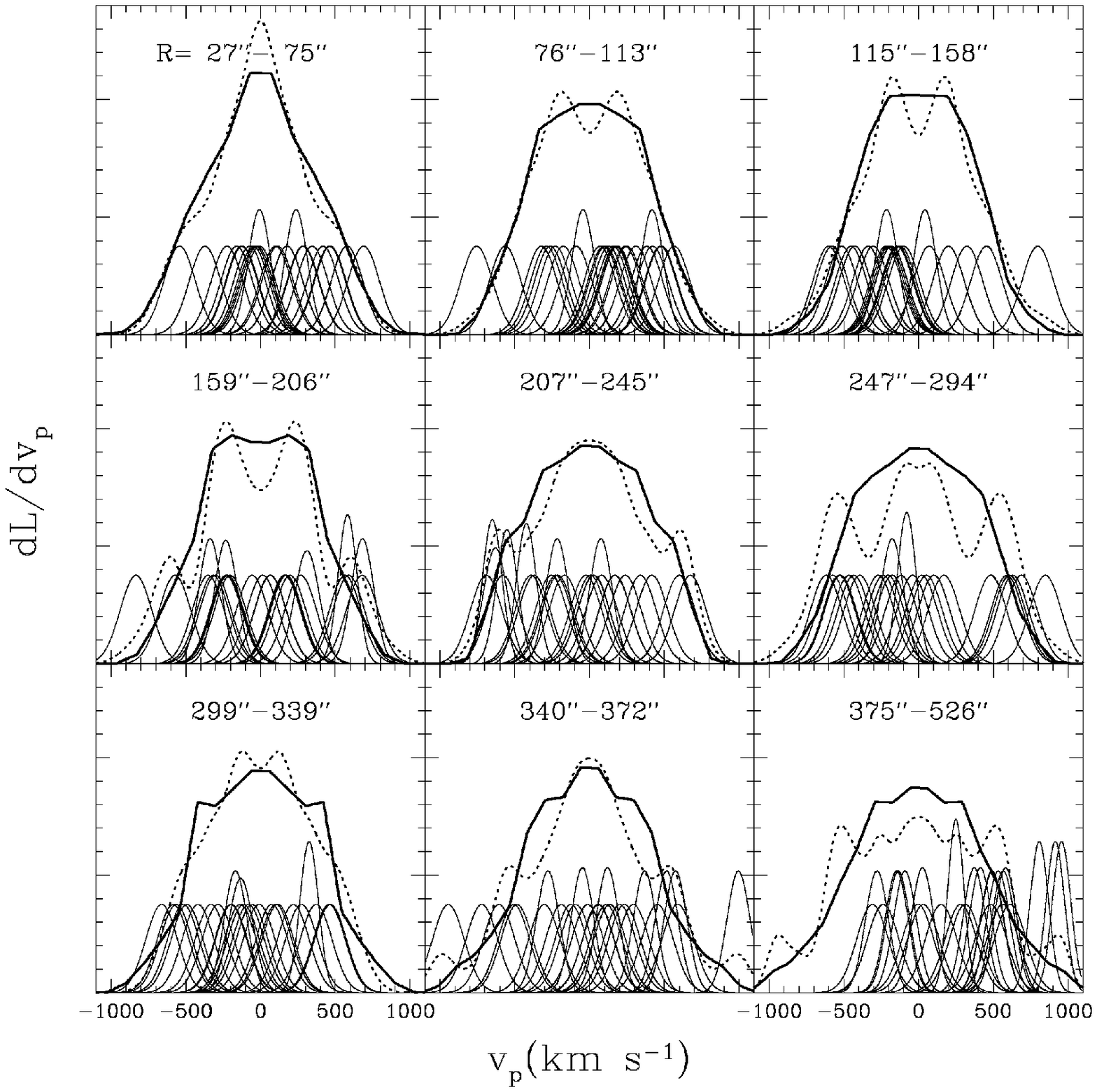}{6.3in}{0}{88}{88}{-47}{-157}}

{\center \small {\bf F{\scriptsize \bf IG}. 7.---}
Line-of-sight velocity distributions for the M87 globular cluster system,
in nine radial bins.
The heavy solid lines show the LOSVDs of the best-fit singular 
isothermal solution 
($\sigma_0=409$ km s$^{-1}$),
averaged in each bin.
The light solid lines show the data (26 points in each bin).
The dotted lines show simulated LOSVDs derived from
the superposition of the data in each bin, symmetrized about $v_{\rm p}=0$.
Compare Fig. 3.
}

{\plotfiddle{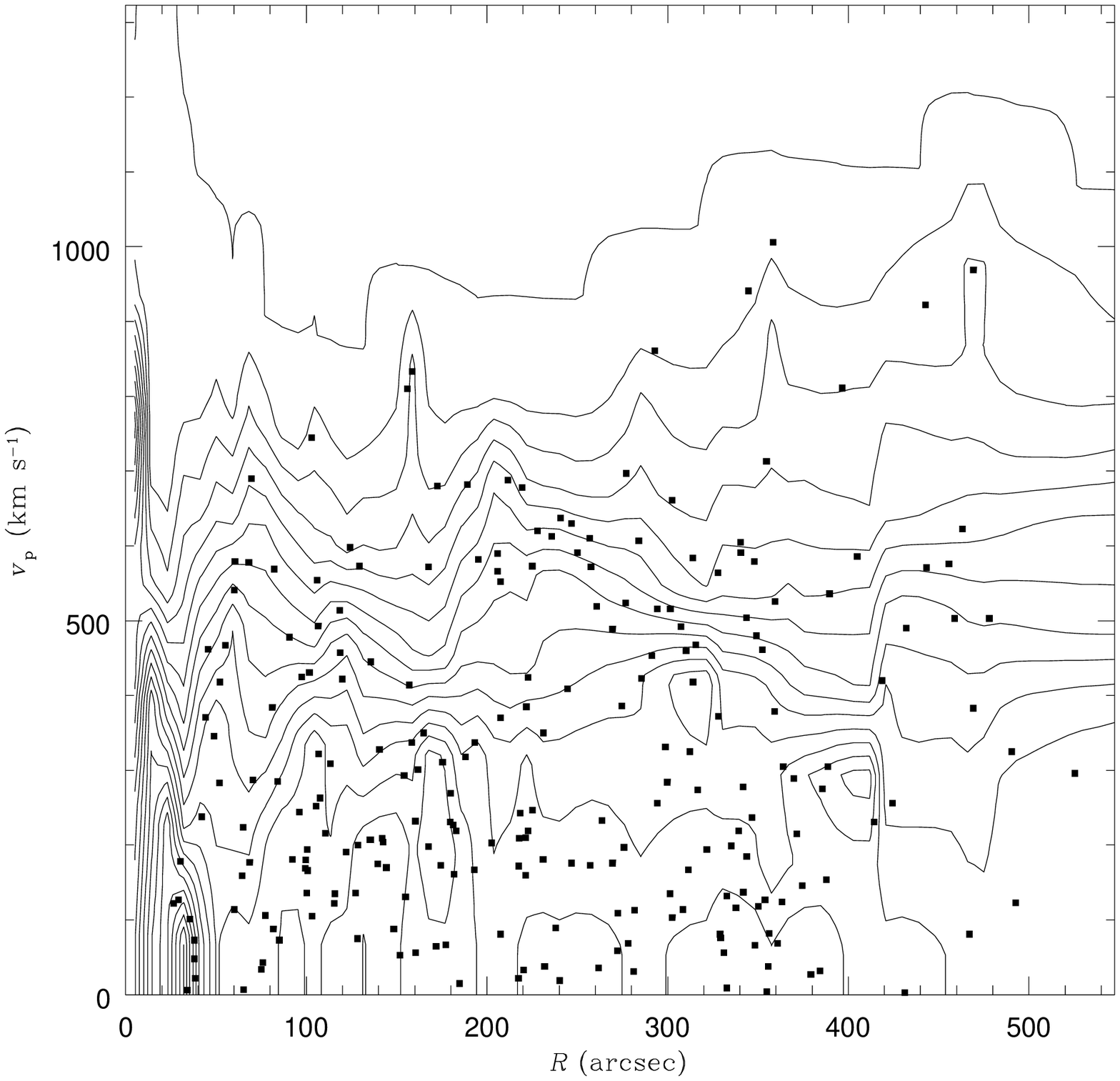}{6.8in}{0}{88}{88}{-35}{-145}}

{\center \small {\bf F{\scriptsize \bf IG}. 8.---}
Contour plot of line-of-sight velocity distribution ($I^{-1} dI/dv_{\rm p}$)
for the best-fit singular isothermal globular cluster system model
($\sigma_0=409$ km s$^{-1}$),
normalized by the surface brightness at each radius $R$.
The last contour represents zero intensity.
The data are shown as squares.
Note that the model LOSVD is peaked at small radii ($R < 50\arcs$)
and is flat and even double-peaked elsewhere.
}

\newpage

\subsection{Generalized Mass Models}

With a constant $M/L$ model ruled out, 
the next question is what the actual radial distribution of mass is
in the system.
The simple SIS model is seen to be inaccurate:
the $\sigma_0$ mass parameters derived separately from the stellar and
GC constraints are inconsistent at the 99\% confidence level
(see Fig. 4 and Table 1).
That the GC-derived mass is {\it higher} than the stellar-derived mass
implies that the radial mass profile $M(r)$ rises more steeply than 
isothermal ($\propto r$) in the outer parts.
To make an initial estimation of $M(r)$, we use 
the spherical isotropic Jeans equations
(see Sargent {\it et al.} 1978 equation 11 and compare CR Fig. 6),
modeling the stellar and GC subsystems separately.
The stellar constraints imply $M \sim r^{0.9}$
out to $r \sim 150\arcs \sim 10$ kpc,
while the GC constraints imply $M \sim r^{1.3}$
outside this radius
(see Fig. 9).
These crude Jeans models thus suggest that it would be fruitful to further explore
two-component mass models consisting of a constant $M/L$
galaxy plus a dark halo with $M(r)$ steeper than isothermal in the
region of $r \sim 100\arcs$--$500\arcs$ (7--36 kpc).

{\plotfiddle{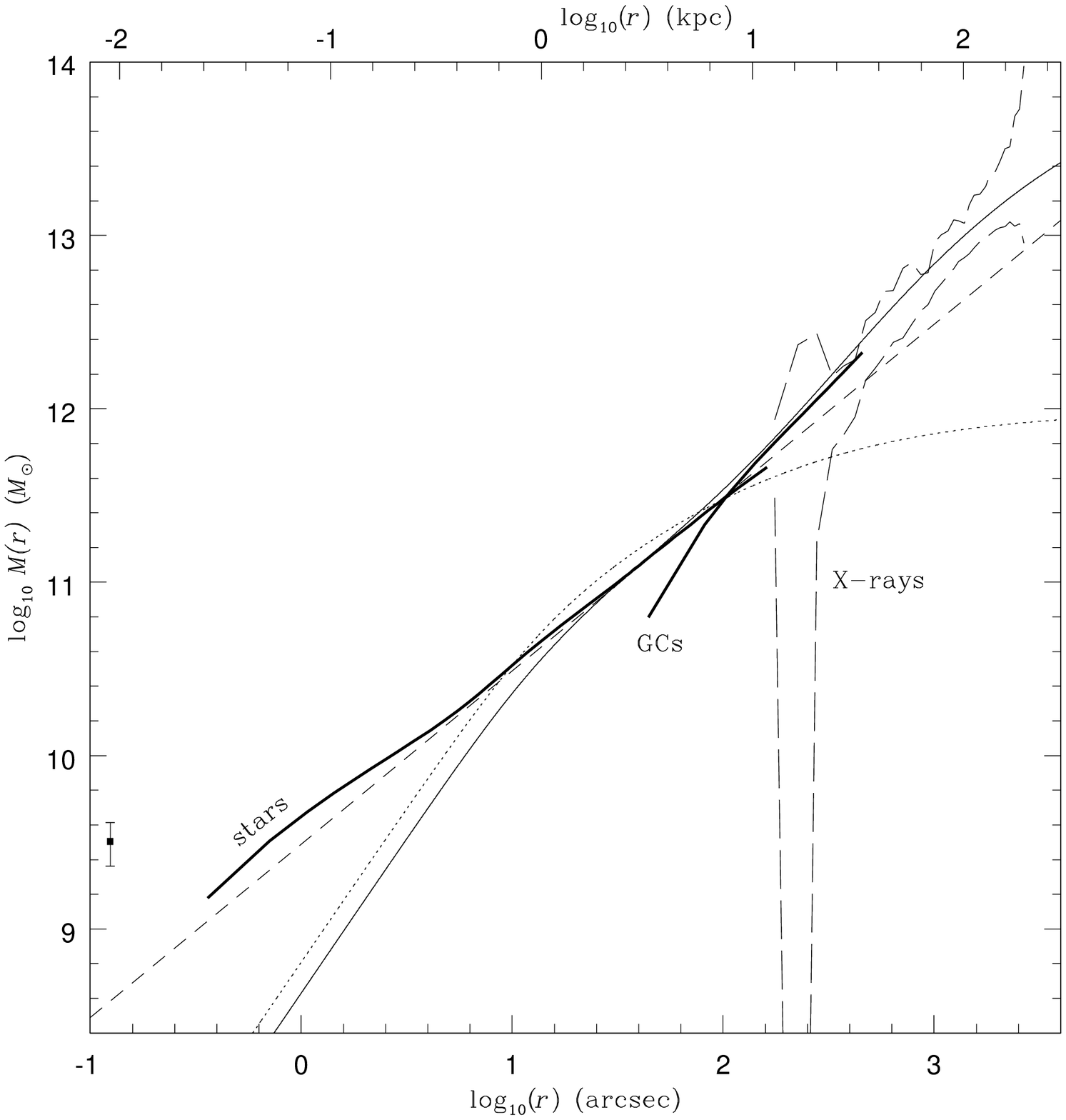}{3.9in}{0}{54}{54}{60}{-100}}

{\center \small {\bf F{\scriptsize \bf IG}. 9.---}
The radial mass profile of M87.
The {\it heavy solid lines} show estimates from the Jeans
equations using the stars and globular clusters separately.
The {\it long-dashed lines} show
confidence limits from the X-ray analysis of Nulsen \& B\"{o}hringer (1995).
The other lines show the best orbit models fit to the combined 
star and GC data, 
for the constant $M/L$ model ({\it dotted}),
the SIS model ({\it dashed}),
and the NFW2 model ({\it light solid}).
The point with error bars shows the estimated mass of the central supermassive
black hole (Macchetto {\it et al.} 1997).
}

\newpage

We next use our orbit modeling methods to
examine such a generalized mass model, where the form of the dark
halo is based on the
the predictions from simulations of cosmologically-motivated structure formation 
({\it e.g.}, Navarro, Frenk, \& White 1996).
This model has a total radial mass density profile of
\[
\rho(r) = \Upsilon_B \, \nu_\star(r) + \frac{\rho_0 r_s^3}{r (r+r_s)^2} ,
\]
where $r_s$ is the break radius between the halo's inner $r^{-1}$ profile
and outer $r^{-3}$ profile.
We construct mass models with a variety of the
parameters $(\Upsilon_B, \rho_0, r_s)$,
fitting the combined stellar and GC constraints,
and report the results of our three best models in Table 2
(see also Figs. 5 and 6).
Note that each distinct model begins with a set of kernels calculated
for two fixed parameters
$(\Upsilon_B/\rho_0, r_s)$,
and may then have its overall mass normalization scaled linearly
to best fit the data.

\begin{tabular}{lccccc}
\\
\hline\hline\\
&
\multicolumn{4}{c}{Model Parameters} &  \\
\cline {2-5} \\
Model name &
\multicolumn{1}{c}{$\Upsilon_B$ ($M_\odot/L_{\odot,B}$)} &
\multicolumn{1}{c}{$\sigma_0$ (km s$^{-1}$)} &
\multicolumn{1}{c}{$\rho_0 r_s^3$ ($10^{12} M_\odot$)} &
\multicolumn{1}{c}{$r_s$ ({}\arcs ; kpc)} &
$\Delta \ln {\cal L}$
\\
\hline \\
\vspace{4pt}
Constant $M/L$ & 9.3 & & & & 208.3 (20.3 $\sigma$)\\
\vspace{4pt}
SIS & & 301 & & & 3.7 (2.3 $\sigma$)\\
\vspace{4pt}
NFW1 & 5.6 & & 0.9 & 500 ; 36 & 0.4 (0.8 $\sigma$) \\
\vspace{4pt}
NFW2 & 5.9 & & 2.4 & 939.3 ; 68 & --- \\
\vspace{4pt}
NFW3 & 6.5 & & 4.8 & 1800 ; 131 & 0.6 (0.9 $\sigma$) \\ \\
\hline
\end{tabular}
{\center \small {\bf T{\scriptsize \bf ABLE} 2.---}
M87 model solutions, given combined stellar and globular cluster 
constraints.
The best-fit model parameters are given for each model
(see text for description).
The difference in log-likelihood between each model and the best
overall model (NFW2) is shown,
along with the statistical significance at which the model is
dispreferred.
}

\vskip 0.4cm

All these models (NFW1, NFW2, and NFW3)
can produce better fits to the full data set than can the best 
SIS model (at the 98\% significance level),
and thus probably better approximate the overall mass distribution.
The mass profiles of these models are consistent with $M(r)$
derived from the galaxy's X-ray halo (Nulsen \& B\"{o}hringer 1995),
both in mass normalization and in density exponent (see Fig. 9).
Both our NFW models and the X-ray model have density profiles that
decline more gradually than isothermal in the galaxy's outer parts
({\it e.g.}, $\rho \sim r^{-1.5}$ at $r \sim 200\arcs \sim$ 15 kpc),
giving a circular velocity curve $v_{\rm c}(r)$ that continues rising
at large radii (Fig. 10).
Most elliptical galaxies have a constant or declining $v_{\rm c}(r)$
at these radii (Gerhard {\it et al.} 2001;
the notable exception is the Fornax cD galaxy NGC 1399),
so this is strong evidence that the dark matter at large radii is
associated with the Virgo Cluster itself.

{\plotfiddle{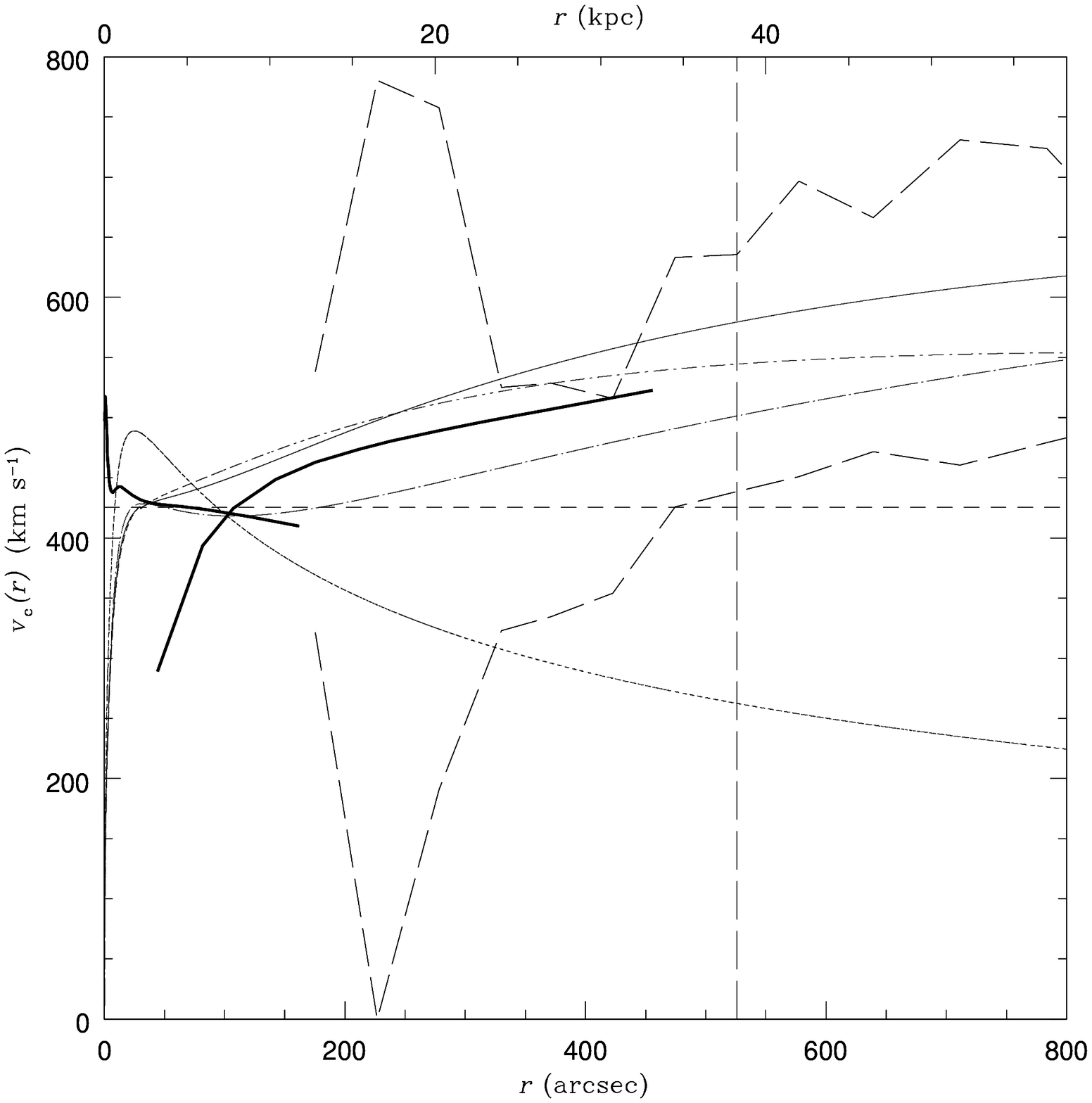}{3.9in}{0}{54}{54}{60}{-100}}

{\center \small {\bf F{\scriptsize \bf IG}. 10.---}
Circular velocity profile of M87,
for different models.
Lines styles are as in Fig. 9,
with the addition of the NFW1 ({\it dot-short dashed})
and NFW3 ({\it dot-long dashed})
models.
The vertical dashed line indicates the outer radius of the
region constrained by the GC data.
}

A caveat about these new models is that
the mass normalization from the stellar and GC data separately
is again inconsistent (at the 1--2~$\sigma$ level).
This may indicate that a still more accurate mass model $M(r)$ could be
found to fit all the existing constraints 
(possibly one with a larger, less-concentrated cluster halo as in McLaughlin 1995),
and/or that one or both of the dynamical subsystems
(stars and GCs)
would be better modeled without the assumption of spherical symmetry.
Further exploration of more detailed models would be useful, but
outside the scope of this project.

\subsection{Orbit Anisotropies}

We have not implemented any method of rigorously exploring the
range of DF functional forms permitted by the data,
so the following results are meant to be suggestive rather than
conclusive.
Fig. 11 illustrates the stellar orbital characteristics of the 
overall best-fit solutions,
where the mean anisotropy is characterized by the parameter
$\beta \equiv 1 - \langle v^2_{\theta}\rangle / \langle v^2_r \rangle$.
The NFW solutions have
a somewhat radial DF in the region constrained by the kinematical data 
($r \sim 0\parcs15$--$170\arcs \sim 0.01$--12 kpc).
The downwards dive in $\beta$
near $r=0\parcs7$ is due to the sharp turnover
in $\hat{\sigma}_{\rm p}(R)$ inside that radius.
Real galaxies presumably have smooth $\beta(r)$ profiles, 
but with no smoothness imposed, our models are somewhat noisy because of the
ill-conditioning of this inversion problem.
Also, the region $r \lsim 1\arcs$ is within the radius of influence
of the central supermassive black hole
(with mass $M_{\rm BH}=3.2\pm0.9 \times 10^9 M_\odot$; Macchetto {\it et al.} 1997),
and thus has not been accurately treated by our models.
In summary, our best models indicate mild radial orbit anisotropy 
($\beta = $ 0.2--0.5)
for the well-constrained stellar region of the galaxy
($r \sim 1\arcs$--$30\arcs \sim $ 0.1--2 kpc)
--- a characteristic which has turned out to be quite typical
for bright elliptical galaxies
(see Gerhard {\it et al.} 2001 and references therein).

{\plotfiddle{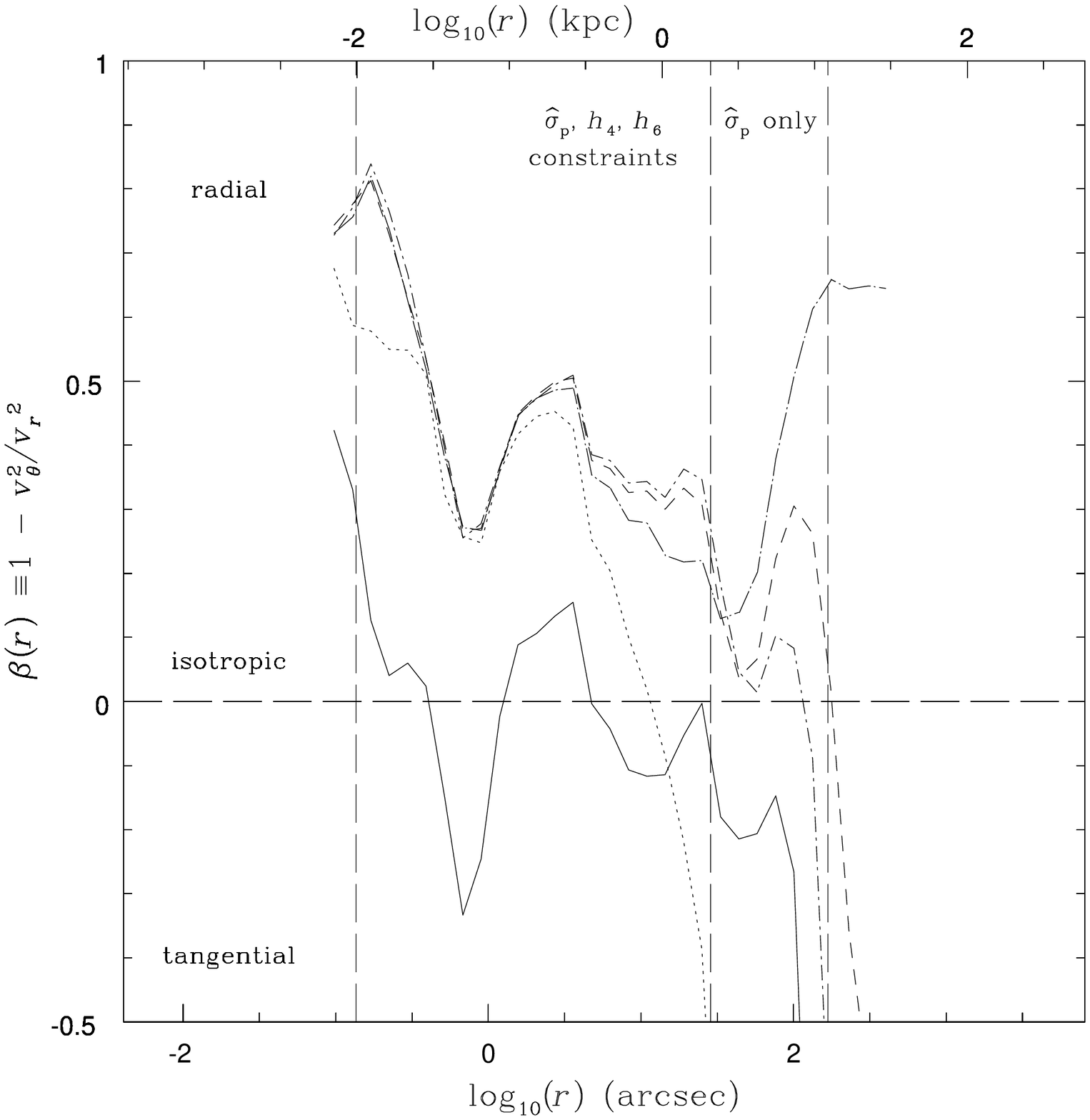}{3.45in}{0}{48}{48}{77}{-90}}

{\center \small {\bf F{\scriptsize \bf IG}. 11.---}
Stellar orbital anisotropy as a function of radius, 
for the overall best-fit solutions from five different mass models:
constant $M/L$ ({\it dotted line}),
SIS ({\it solid line}),
NFW1 ({\it dot-short dashed line}),
NFW2 ({\it dashed line}),
and NFW3 ({\it dot-long dashed line}).
The vertical dashed lines indicate the 
radial regions constrained by the stellar kinematical data.
}

\vskip 0.55cm

Fig. 12 shows the orbital characteristics of the globular cluster system
for the overall best-fit solutions.
There is considerable variation between the best NFW solutions,
so we cannot yet strongly constrain the GC anisotropy,
except to note that it seems roughly isotropic at large radii
($r \sim 100\arcs$--$500\arcs \sim $ 10--40 kpc).
This seems to contradict the tangential anisotropy one would expect
given the flat and double-peaked LOSVDs
seen in the data 
and in the models that fit only the GC data
(see \S 2.4 and Figs. 7 and 8);
--- illustrating further the need for a mass model
that better fits both the stellar and GC constraints.
Note that the solutions' radial anisotropy at $r \lsim 40\arcs \lsim $ 3 kpc should not be considered
highly significant since there are only 8 GC velocities measured inside
that projected radius.

{\plotfiddle{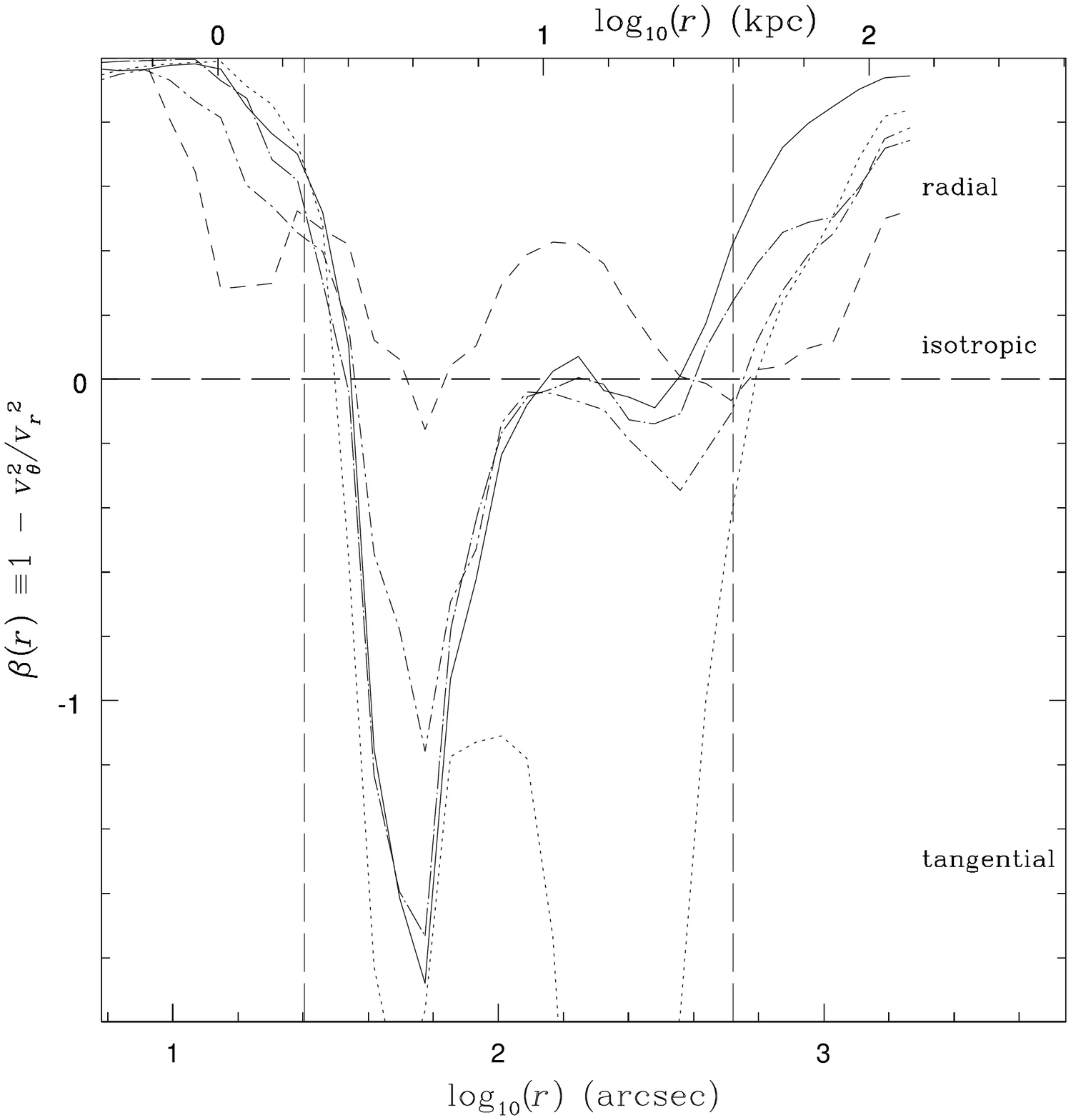}{4.2in}{0}{55}{55}{60}{-93}}

{\center \small {\bf F{\scriptsize \bf IG}. 12.---}
Velocity anisotropy of globular cluster system
as a function of radius,
for different mass models
(see Fig. 11 for line descriptions).
The vertical dashed lines indicate the 
radial region constrained by
the GC velocity data.
}

\section{CONCLUSIONS}

We have developed a very general spherical orbit modeling method that makes
full use of all the information in a set of discrete velocity data,
and have applied it to the giant elliptical galaxy M87 using all the available dynamical
constraints
(measurements of the spatial distribution and the kinematics of the stars and 
of the globular clusters).
By comparing two simple mass models ---
a constant mass-to-light ratio galaxy and a singular isothermal sphere ---
we find that a constant $M/L$ is ruled out to a high level of significance.

We further implement some more generalized mass models in order to more
accurately represent the galaxy as consisting of a constant $M/L$
stellar system plus a dark halo.
We find improved solutions which have a dark halo density profile
decreasing more slowly than $\sim r^{-2}$
at large radii (10--40 kpc).
This very extended halo seems best interpreted as belonging to
the Virgo Cluster itself, rather than to M87.

There is substantial room for improvement of our M87 models.
First, measurements of higher-order velocity moments for the stars
are available to only $\sim 0.3 R_{\rm eff}$.
Such measurements at larger radii would be quite helpful in better
constraining the models,
and can be most efficiently obtained by measuring
discrete velocities of the galaxy's planetary nebulae.
Second, a larger range of mass models could be explored to find
solutions that fit the stellar and GC data
together more consistently.
And third, future models could relax our assumption of spherical symmetry,
providing a more accurate model of the system in its outer parts, 
where flattening is seen in both the stellar and GC systems.


The same techniques should also be applied to other, non-cD giant elliptical galaxies
in clusters (such as NGC 4472), as well as to galaxies in the field.
The different formational histories of these galaxies (via mergers, accretion, etc.)
may result in significantly different
mass distributions and dynamics.
Furthermore, the colors, abundances, and spatial distributions of the GCs in M87 
and other galaxies
indicate that they comprise two or more distinct populations
({\it e.g.}, Neilsen \& Tsvetanov 1999; Gebhardt \& Kissler-Patig 1999).
It would be interesting to use our orbit modeling methods to
look for any dynamical differences between these populations, which
may shed light on their formation histories,
for which several different scenarios have been proposed
({\it e.g.}, Ashman \& Zepf 1992; Forbes, Brodie, \& Grillmair 1997; Harris, Harris, \& McLaughlin 1998; C\^{o}t\'{e}, Marzke, \& West 1998).
The available set of kinematical tracers observed in the outer parts of early-type galaxies
is now increasing rapidly, and we are hopeful that a much clearer picture of galaxy halos
is forthcoming.

\vskip 67pt

We thank
John Blakeslee, Judy Cohen, Dan Fabricant, Marijn Franx, Lars Hernquist,
Konrad Kuijken, and the anonymous referee for helpful comments;
Ken Sembach for providing his data in tabular form;
John Huchra, Roeland van der Marel, and Dean McLaughlin for both;
Arunav Kundu for providing his paper in advance of publication,
and his data in tabular form.
C.S.K. is supported by the Smithsonian Astrophysical Observatory.

\end{document}